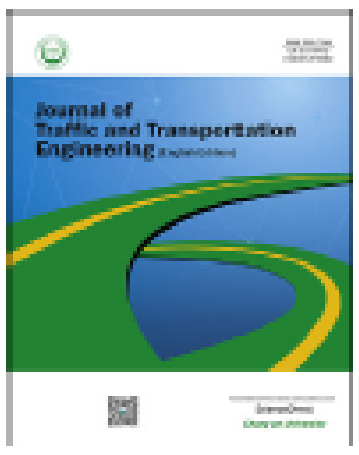

Original research paper

# Investigating Short-term and Long-term Binder Performance of High-RAP Mixtures Containing Waste Cooking Oil

Hamed Majidifard[1*], Nader Tabatabaee[2], William Buttlar[3]

[1*]Corresponding author, Graduate Research Assistant, Department of Civil & Environmental Engineering, University of Missouri, Columbia, Missouri, 65211, Email: hamed.majidifard@mail.missouri.edu

[2] Professor, Department of Civil Engineering, Sharif University of Technology, Tehran, Iran, Email: nader@sharif.edu

[3]Professor, Glen Barton Chair in Flexible Pavements, Department of Civil & Environmental Engineering, University of Missouri, Columbia, Missouri, 65211,

Email: buttlarw@missouri.edu

**Highlights**

- The use of waste cooking oil as a recycling agent opens the possibility for the routine design of 60-to-nearly-100% recycled content asphalt paving mixtures.
- Aging susceptibility of recycled binders with waste cooking oil is higher than virgin binder.
- Oil has a greater effect on reducing the stiffness of RAP binder than increasing its m-value.
- Waste cooking oil tended to improve mixture workability and low temperature performance while reducing moisture and rutting resistance.
- Selecting the optimum oil content as equal to the average oil content based on satisfying the LT and HT PG can assure short-term and long-term performance.

**Abstract**

The environmental and economic benefits of recycling asphalt pavements have received much attention in recent years. Because of the increase in the cost of raw materials and energy carriers, the reuse of large portions of reclaimed asphalt pavement (RAP) is critical in reducing both the cost and environmental footprint of asphalt pavements. High-RAP mixtures are more prone to low temperature cracking and poor mixture workability because of the higher stiffness of RAP binder. Recycling agents are one of the additives which are used to improve these deficiencies. However, there is some ambiguity about the optimum content of recycling agent to assure proper performance of recycled asphalt pavement during its service life. The current study used 60% and 100% fractionated RAP with waste cooking oil as a recycling agent and crumb rubber to alleviate the aforementioned problems. Laboratory evaluation showed that increasing the amount of recycling agent in the high-RAP mixtures improved their workability and low temperature performance while decreasing moisture damage and rutting resistance. The long-term susceptibility to aging of recycled binder with the organically-based recycling agent was also investigated. A procedure to obtain the optimum percentage of recycling agent was devised to strike a balance between the performance characteristics of mixtures with a high-RAP content.

## 1 Introduction

The use of RAP in hot-mix asphalt (HMA) decreases the construction cost, reduces material transportation costs and promotes sustainability (Wang et al., 2018). RAP contains a higher amount of fine aggregate than the original HMA as a result of milling. This problem can be addressed by fractionation of the RAP into different sizes. Aged binder in RAP is stiffer, which yields mixes with poor low temperature (LT) cracking and workability characteristics (Buttlar et al., 2018a; Majidifard et al., 2019). To rectify these deficiencies, researchers and practitioners have tried methods such as the use of softer fresh bitumen, warm mix asphalt (WMA) additives, foamed bitumen and recycling agents (Buttlar et al., 2018b; Jahangiri et al., 2019).

Generally speaking, all of these methods can be considered as recycling agents. More specifically, recycling agents act as either a softener to reduce the effect of the higher stiffness of the aged binder and improve the workability or as rejuvenators to reduce the stiffness of the aged binder by reversing the process of aging by introducing some of the lost components of the aged asphalt to improve workability and reduce the cracking potential (Tabaković et al., 2017). Depending upon the type of WMA additive used and its role as a lubricant or surfactant, these products may permit a lower mixing temperature, which can in turn inhibit further aging of the RAP binder. Softer virgin binders can be effectively used when the RAP binder is not severely aged. This relies on the premise that some degree of blending of the fresh and RAP binder will result in an appropriate stiffness level for the effective binder system in the mixture.

The use of some waste materials in any of the above processes promotes pavement sustainability. Tahami et al., 2018 concluded that the usage of rice husk ash (RHA) and date seed ash (DSA) fillers improved the rutting and fatigue performance of asphalt mixtures compared to control ones (Tahami et al., 2018). Seidel and Haddock found that soy fatty acids can serve as a fluxing agent to enhance the properties of recycled asphalt binders (Seidel and Haddock, 2012). Generally, bio-based oils are highly susceptible to aging, owing to the high percentage of volatile components in their chemical make-up (Golalipour, 2013; Yang and You, 2015). However, waste cooking oil is supposed to have lower aging susceptibility in comparison to fresh bio-oils due to undergoing the cooking process at high temperature which leads to significant reduction of volatile components (Chen et al., 2014). Furthermore, the flash point of waste cooking oil is above 220 ºC which means that waste cooking oil is stable enough for application in hot mixing asphalt mixtures (Chen et al., 2014). An ample supply of waste cooking oil exists, with roughly 5 million tons generated in China alone each year (Chen et al., 2014). Several research studies and industrial activities have further positioned the feasibility of using waste cooking oil as a rejuvenator in recent years (Chen et al., 2014; Asli et al., 2012; Bailey and Phillips, 2010).

The results of chromatography and fourier-transform infrared spectroscopy (FTIR) testing reveal that aging decreases the asphaltene content of binder and increases the intensity of carbonyl bonds in the aged binder, respectively (Elkashef et al., 2017; Yang et al., 2014). Chromatography shows that the addition of waste vegetable oil to aged asphalt binder decreases the asphaltene content (Chen et al., 2014; Asli et al., 2012). FTIR tests have revealed that the intensity of the carbonyl and sulfoxide bonds in aged asphalt decreases when waste vegetable oil is added (Chen et al., 2014; Gong et al., 2015).

The bio-oils in these studies exhibited the characteristics of a rejuvenator. The rate of aging in a recycled, aged binder in the presence of various recycling agents upon re-placement in service (i.e., 'resumed aging' or 're-aging') is very complex, and very important (Mohammadafzali et al., 2015). The rate of re-aging in the long run will obviously depend on the type of recycling agent used. Compared to the rate of aging in fresh binders, the use of crude, water-based emulsions and heavy paraffinic extract oil recycling agents were found to reduce the rate of re-aging, whereas the use of petroleum natural distillate and bio-oil fatty acid increased the rate (Mohammadafzali et al., 2015). In another study, Yang et al. concluded that the potential of aging of bio-modified binders is higher than virgin binder according to RV and DSR testing (Yang et al., 2014). A recent study showed that after 40 hours of PAV aging to simulate long-term aging, the measured effect of rejuvenation had been minimized (Cucalon et al., 2017).

The addition of recycling agents to asphalt binder has been shown to improve its low temperature properties, but may reduce rutting resistance (Chen et al., 2014, Mogawer et al., 2017; Jahangiri et al., 2018); The results of penetration tests on recycled aged binder with conventional petroleum and novel organic recycling agents, including organic oil, aromatic extract, waste engine oil, distilled tall oil, waste vegetable oil and waste vegetable grease, showed that all of the recycling agents increased the penetration of aged binders from 19 to 78 dmm, which was very similar to the penetration of the fresh binder (Zaumanis et al., 2013). Mirhosseini et al. (2018) used Rotational Viscometer (RV) and DSR test to examine the application of date seed oil (DSO) as a rejuvenating agent in RAP-blended mixes. They concluded that adding 10% DSO to the binder decreased viscosities of the blends significantly and deteriorated the rutting resistance of the binder based on DSR testing results. However, fatigue performance was improved according to linear amplitude sweep (LAS) test results (Mirhosseini et al., 2018). Also, Mazzoni et al. (2018) investigated the effect of different rejuvenators on asphalt binder blends. They confirmed that the type and dosage of rejuvenator should be carefully selected considering the virgin and RAP binder properties, since their interaction greatly influences the ultimate mixture performance (Mazzoni et al., 2018).

According to the previous literature, there is still some ambiguity about which criterion (assimilate RAP binder properties to virgin binder properties based on penetration, HTPG[1] or LTPG[2]) can assure proper performance of recycled asphalt pavement during its service life. This study was undertaken to determine the optimum content of the recycling agent to strike a balance between the short and long-term performance characteristics of high-RAP mixtures. Furthermore, in this study, crumb-rubber is used to compensate for the reduction of rutting resistance which caused by adding rejuvenator. On the other hand, using crumb rubber and waste cooking oil improves the sustainability of the pavement (Rath and Buttlar, 2018). The cost of the recycled high-RAP mixture with waste cooking oil is still economic even if crumb rubber is used to improve rutting performance.

## 2 Material characterization

### *2.1 Asphalt binders*

Superpave PG 58-22 was used as the virgin binder in this study. The virgin binder hereafter is called fresh binder with an abbreviation of FB. Bending beam rheometer (BBR) and dynamic shear rheometer (DSR) tests were conducted on the FB in accordance with AASHTO T313-10 and AASHTO T315, respectively. The continuous low temperature (LT) and high temperature (HT) performance grade (PG) of the binder were determined to be -25.5 and 61.1 °C, respectively. The continuous RAP binder PG was obtained as 82.3-10.3 using non-extraction recovery method (Table 3). Also, 12% crumb rubber by weight of virgin binder was considered to investigate the effect of crumb rubber in high-RAP mixtures. Crumb rubber was added to the virgin binder at 160°C for 2 hours (5000RPM) in a high shear mixer. Afterward, the modified crumb rubber binder was added to RAP and virgin aggregates.

[1] High Temperature Performance Grade
[2] Low Temperature Performance Grade

### *2.2 Recycling agent*

Waste cooking oil was used as the recycling agent in this study. Fatty acids constituted most of the cooking oil structure used. The oil had a specific gravity of 0.941 g/cm$^3$ and a viscosity of 19.5 cSt at 60°C. Furthermore, the chemical properties of the recycling agent are shown in Table 1.

**Table 1.** Chemical properties of waste cooking oil

| Fatty Acid Type | Fatty Acids' Formulation | Wt(%) | Saturation Type |
|---|---|---|---|
| Oleic acid | C18:1 (Cis 9) | 39.8 | Unsaturated |
| Palmitic acid | C16:0 | 28.4 | Saturated |
| Linoleic acid | C18:2 (Cis) | 25 | Unsaturated |
| Stearic acid | C18:0 | 4.2 | Saturated |
| α-Linolenic acid | C18:3 alpha | 0.8 | Unsaturated |
| Myristic acid | C14:0 | 0.63 | Saturated |
| Vaccenic acid | C18:1 t | 0.3 | Unsaturated |
| Palmitoleic acid | C16:1 | 0.2 | Unsaturated |
| Linoleic acid | C18:2 t | 0.17 | Unsaturated |
| Lauric acid | C12:0 | 0.11 | Saturated |
| Lignoceric acid | C24:0 | 0.09 | Saturated |
| Margaric acid | C17:0 | 0.04 | Saturated |
| Pentadecylic acid | C15:0 | 0.03 | Saturated |

### *2.3 Aggregates*

The aggregates used in both virgin and RAP mixes were considered to be siliceous. X-ray fluorescence (XRF) of the virgin and RAP aggregates indicated a composition of 55% and 53% of $SiO_2$, 16.8% and 12% of CaO, and 8.5% and 10.4% of $Al_2O_3$, respectively.

## 3 Experimental setup and procedure

Fig 1 depicts the flow chart for short- and long-term performance testing on recovered and artificial RAP binder.

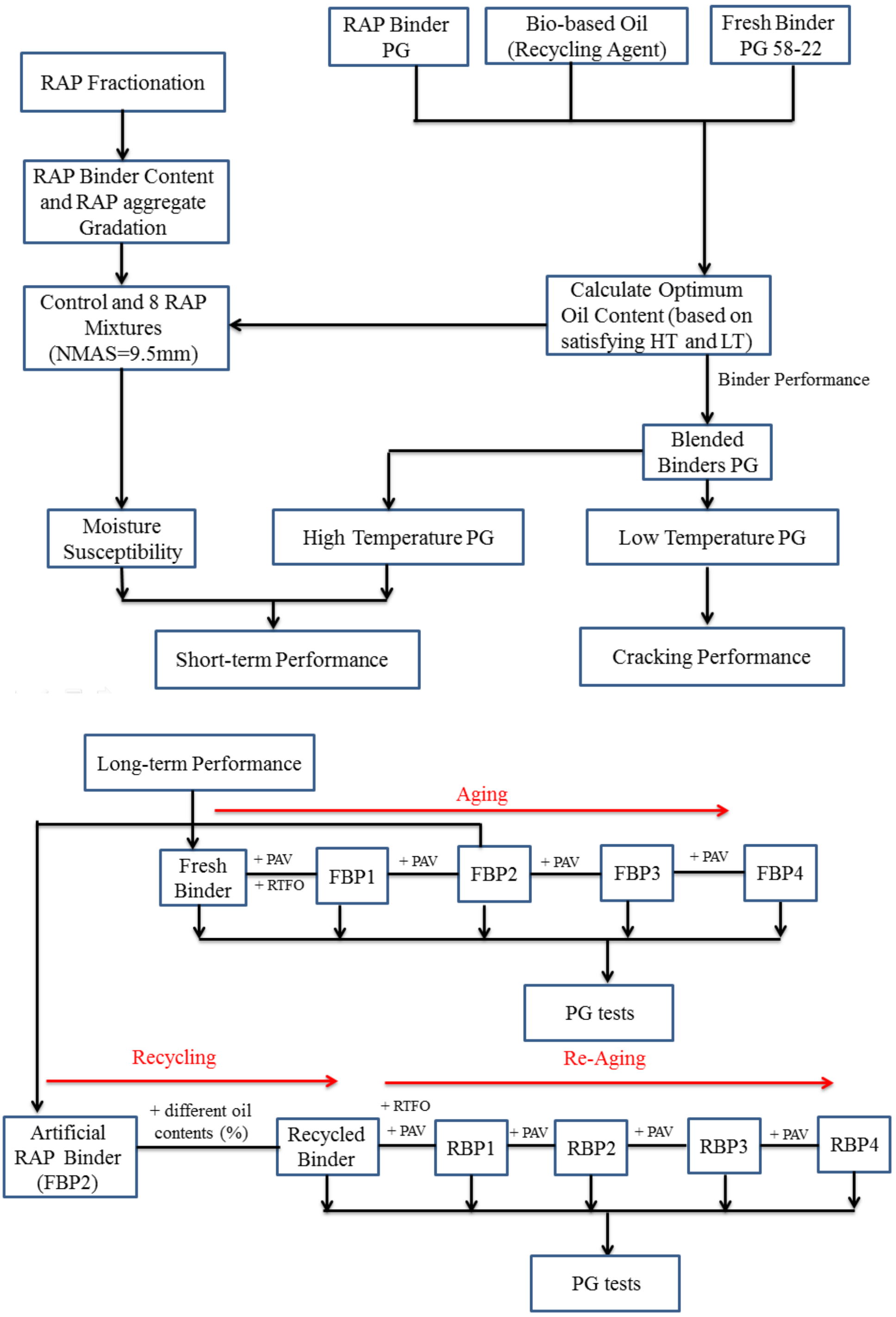

**Fig 1**. Experimental plan.

The testing procedure in this study consisted of the following steps:

Section 1:

1. Determine the PG of the RAP binder
2. Determine the optimum oil content to be added to the RAP to closely match the PG of the FB

3. Prepare high-RAP HMA and find the optimum binder content
4. Determine the PG of the blended binders in accordance with the RAP binder replacement (RBR), virgin binder, and oil
5. Perform moisture susceptibility testing

Section 2:

1. Use rolling thin film oven (RTFO) and 4 pressure aging vessel (PAV) cycles on fresh binder
2. Select 2PAV binder as the artificial RAP binder (FBP2)
3. Recycle the FBP2 by adding different oil contents
4. Determine the optimum oil content to add to the FBP2 to closely match the PG of the FB
5. Re-age the recycled binder
6. Determine the PG of the recycled binder

### *3.1 Determination of RAP binder PG, blended binder PG and optimum oil content*

The non-solvent extraction method was used to determine the RAP binder PG (Bahia et al., 2011). Because the RAP binder was not recovered in this research, there was no possibility of adding oil to it in order to calculate the PG temperatures of RAP binder + oil. The procedure used by Teymourpour was used in this study. The oil and fresh binder was synthesized in a high shear mixer (5000 rpm) at 150°C for 30 min based on their percentage (wt.) in the final blend for each mixture (Teymourpour et al., 2015) (Table 2). RTFO and PAV aging were performed on binders in accordance with AASHTO 240-09 and ASTM D6521, respectively. The binders used (fresh, fresh + crumb rubber, fresh + oil and fresh + crumb + oil) were tested in the unaged and RTFO-aged conditions using a dynamic shear rheometer (DSR) according to AASHTO T315-10 and under PAV-aged conditions using the bending beam rheometer (BBR) (AASHTO T313-10) to determine their performance grade according to AASHTO 29-08. RAP binder PG was calculated according to the non-solvent method (Bahia et al. 2011). The continuous grade of the RAP binder was found to be PG 82.3-10.3. The blended binder performance grade was calculated as shown in Equation (1) (McDaniel and Anderson, 2001) based on the grades and percentages of constituents as shown in Table 2.

$$T_{blend} = \%RBR \times T_{RAP} + (1 - \%RBR) \times T_{fresh} \quad (1)$$

Where $T_{blend}\%RAP \times T_{RAP} + (1\text{-}\%RAP) \times T_{virgin}T_{(blend)}$ is blended binder PG, $T_{(RAP)}$ is RAP binder PG, $T_{(fresh)}$ is the fresh or (fresh + oil) binder PG and RBR is the RAP binder replacement.

Equation (1) was used to calculate $T_{RAP+oil}$. The HT and LT PG of the RAP binders with different oil contents were then determined. As expected, the HT and LT PG decreased as the oil content increased. Fig 2 shows that adding 10% and 16% of bio-oil to the RAP binder with a HT of 82.3°C reduced the HT of the RAP to 68.7°C and 61.1°C, respectively. Conversely, these additions, improved the LT of the RAP from -10.3°C to -22°C and -28.2°C, respectively. Both of these oil contents brought the HT and LT of the blends to levels that easily met or exceeded those of PG58-22. From the economic point of view, it is preferable to add as much oil as needed, and to carry out fine-tuning as needed. The results show that by adding a proper amount of bio-oil to the aged binder, the rheological properties reverted to a state similar to the original binder. Although chemical tests were not performed in this research, similar studies have demonstrated that adding waste vegetable oil to aged binder restores some of the chemical properties of the aged binder (Chen et al., 2014; Asli et al., 2012; Gong et al., 2015)

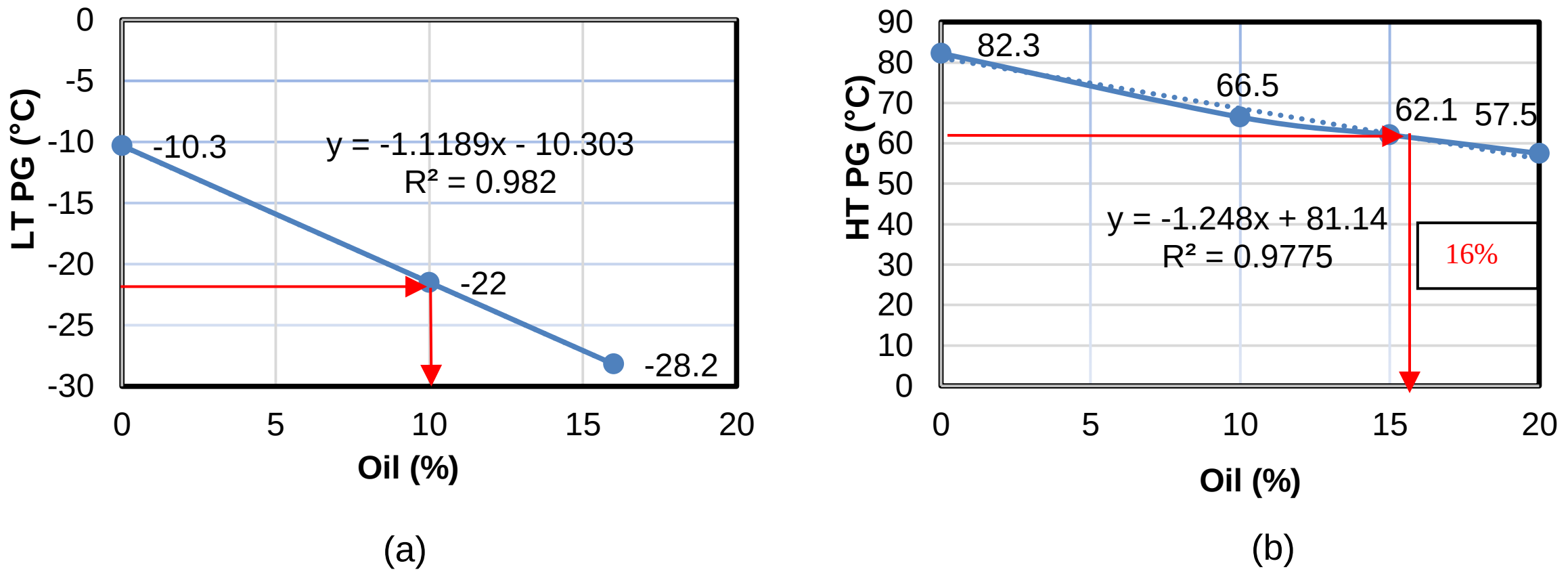

**Fig 2.** Increase in oil content and decrease in: (a) LT PG; (b) HT PG.

The best performance for LT PG was the binder with 16% oil, while the binder with crumb rubber showed the best HT PG (Table 2). Interpolation between the percentage of oil and continuous PG of the two tested binders indicated that the addition of 13% oil yielded a continuous grade of PG 63.2-25.1, which met the properties of the fresh binder. Note that the 60% RAP + CRM + 16% O mixture has one of the highest useful temperature interval (UTI) ranges of all the combinations (Table 2). This result supported the use of crumb rubber along with the recycling agents in high-RAP mixtures.

Thermal cracking is one of the prevalent failure modes of asphalt pavements in the cold regions (Behnia et al., 2011; Dave et al., 2013; Keshavarzi and Kim, 2016; Wagoner et al., 2005). In this research, improving the low temperature performance of the mixture was of utmost importance. Thus, the 16% of oil which was obtained to satisfy the HT PG produced the best LT PG (-27 for 60% RAP and -27.9 for 100% RAP) of all mixtures. Their nominal HT PG was determined to be 58°C grade.

**Table 2.** Blended binder PG for various mixtures

| **Mixture Type** | **RBR** | **TB** | **Continuous HT** | | **Nominal PG** | | **Continuous UTI±** |
|---|---|---|---|---|---|---|---|
| | (%) | (%) | LT (°C) | HT (°C) | LT (°C) | HT (°C) | (°C ) |
| 0% RAP | 0 | 5.9 | 61.1 | -25.5 | 58 | -22 | 86.6 |
| 60% RAP+FB (47.8% + 52.2%) | 47.8 | 7.3 | 71.3 | -18.2 | 70 | -16 | 89.5 |
| 60% RAP+FB+16% O* (47.8%+44.6%+7.6%) | 47.8 | 7.3 | 61.1 | -27 | 58 | -22 | 88.1 |
| 60% RAP+FB+13% O (Int) (47.8%+46%+6.2%) | 47.8 | 7.3 | 63.2 | -25.1 | 58 | -22 | 88.3 |
| 60% RAP+FB+10% O** (47.8%+47.4%+4.8%) | 47.8 | 7.3 | 65.1 | -23.4 | 58 | -22 | 88.5 |
| 60% RAP+CRM (45.1%+54.9%) | 45.1 | 7.7 | 78 | -18.9 | 76 | -16 | 96.9 |
| 60%RAP+CRM+16%O (45%+48%+7%) | 45 | 7.7 | 63.5 | -27.4 | 58 | -22 | 90.9 |
| 100% RAP+FB+16% O* (76.1%+11.7%+12.2) | 76.1 | 7.6 | 61.1 | -27.9 | 58 | -22 | 89 |
| 100% RAP+FB+13% O (Int) (76.1%+14%+9.9%) | 76.1 | 7.6 | 64.4 | -24.9 | 64 | -22 | 89.3 |
| 100% RAP+FB+10% O** (76.1%+16.3%+7.6%) | 76.1 | 7.6 | 67.4 | -22.2 | 64 | -22 | 89.6 |

*Oil content was selected based on satisfying HT PG

**Oil content was selected based on satisfying LT PG

±UTI: useful temperature interval

TB = total (binder + oil) = % RB + % FB + % oil

Int = interpolated

| Table 3. Fresh, RAP and artificial RAP binder PG | | | | | |
|---|---|---|---|---|---|
| **Binder Types** | **Continuous HT** | | **Nominal PG** | | **Continuous UTI±** |
| | HT (°C) | LT (°C) | HT (°C) | LT (°C) | (°C ) |
| Fresh Binder | 61.1 | -25.5 | 58 | -22 | 86.6 |
| RAP Binder (in mixture) | 82.3 | -10.3 | 82 | -10 | 92.6 |
| Artificial RAP Binder | 76.7 | -22.7 | 76 | -22 | 99.4 |

*3.2. Aging susceptibility of recycled binder (long-term performance)*

The fresh binder was aged in the RTFO for 85 min at 165°C to simulate short-term aging, followed by PAV aging at 100°C for 20 hours to simulate long-term aging. Up to four cycles of PAV aging were used for evaluation of aging susceptibility. The 2PAV binder was selected to represent the artificial RAP binder (FBP2) (Ma et al., 2017) and different contents of oil were added and mixed with it at 5000 rpm for 30 min using a high-shear mixer. The blended binder was then aged by RTFO and PAV to simulate the long-term aging that a RAP HMA mix would experience (RBP1). Linear regression of the LTPG and percentage of added oil was performed to determine the percentage of oil required to soften the artificial RAP binder to meet the PG grade of the fresh binder to be 2.9% for LT and 9.2% for HT. These optimum oil contents of 9.2% and 2.9 % were different from the optimum oil contents of 16% and 10% in RAP mixtures. Different sources of RAP binder are thought to be the main cause of this discrepancy. The RAP binder grade used to create the study mixtures was PG 82.3-10, while the artificial RAP binder grade was PG 76.7-22.7.

To evaluate recycled binder aging susceptibility in the long-term, the recycled binder was subjected to 1, 2, 3 and 4 PAV cycles to simulate re-aging, and BBR and DSR testing were performed on the specimens after each PAV cycle, as shown in Fig 1.

*3.3. Asphalt mixture preparation*

Following industry practices in Iran, the Marshall mix design method was used for determination of the optimum asphalt binder. The pertinent AASHTO standards were used for sample preparation, compaction and determination of various specific gravities needed. The optimum binder content corresponding to the target air voids of 4% were determined. The results showed that for high-RAP mixtures, the air voids were 5.5% and 5.6%, which fell outside of the typical 3% to 5% limit, but the addition of proper amounts of oil made it possible to meet the 4% target air voids. It should be mentioned that the 60%RAP+FB and 60%RAP+CRM did not meet the target air void criteria; therefore, their results are included here for comparison purposes only.

### *3.4. Moisture susceptibility of mixtures (short-term performance)*

Moisture damage in asphalt mixtures manifests itself as adhesive and cohesive failure. In the former, the intrusion of water leads to the detachment of binder from aggregates. In the latter, the failure occurs in the mastic binder upon the intrusion of water (Kringos et al. 2008; Al-Qadi et al. 2014). For RAP mixtures, the cohesive resistance in mastic binder could be lower than that of the fresh binder (Mogawer et al. 2012). Because of the oleophilic-hydrophilic molecular structure of waste vegetable oil, the oil-modified binders exhibited high moisture susceptibility (Gong et al. 2015). Moisture susceptibility was thus evaluated according to AASHTO T283-11 by conducting indirect tensile (IDT) strength testing on 60 specimens.

## 4 Result and discussion

The test results are reviewed and discussed in this section.

### *4.1 Aging susceptibility of recycled binder (long-term performance)*

Fig 3 shows the results of BBR and DSR testing on artificially-aged binders through four 20-h cycles of PAV aging. Generally, as the binders age, their stiffness at low and high service temperatures increases. Figs 3(a) and 3(b) show the corresponding continuous LT and HT performance grades.

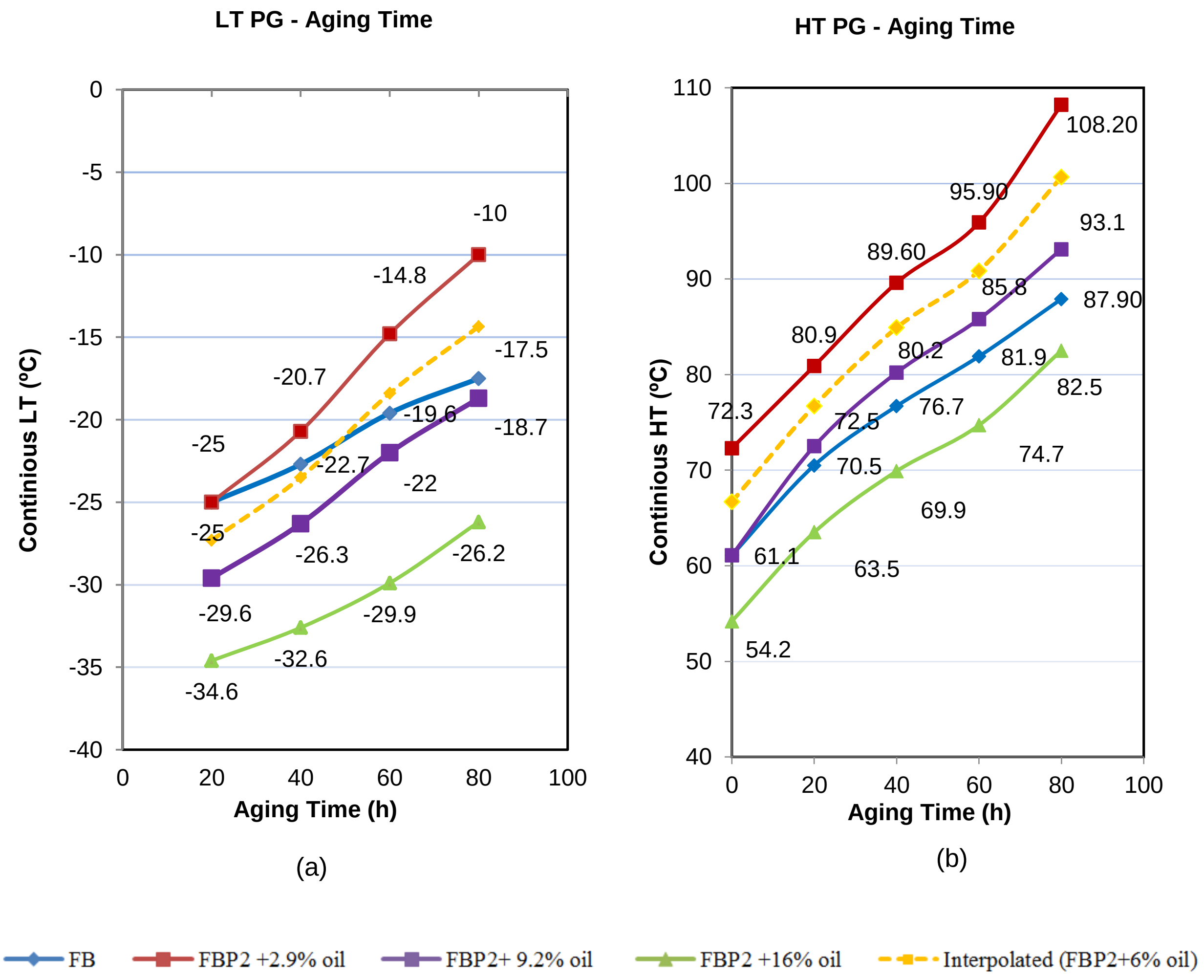

**Fig 3.** Effect of binder aging on: (a) LT PG; (b) HT PG.

Fig 3 shows that the recycled binders aged faster than fresh binders under long-term aging. Aged binder recycled with 2.9% oil had the same LT as the original binder after the first PAV cycle (-25°C). After continuing the PAV cycles, the continuous PG increased faster than for the fresh binder. If -17.5°C is considered to be a failure PG temperature after four PAV cycles (80 h) for the fresh binder, recycled binder with 2.9% oil reached -17.5°C after 51 h, which is 29 h shorter than the time for the fresh binder. This confirms the importance of investigating aging susceptibility of recycled binders in the long-term (Raab et al., 2017).

The results demonstrated the high aging susceptibility of recycled binders with waste cooking oil (Figs 3 and 4); therefore a higher oil content should be considered when recycling the RAP binder to assure its long-term performance. The recycled binder with 9.2% oil which was selected to satisfy the HT PG criterion showed better LT PG than the fresh binder after the first PAV cycle (Fig 3); however, after applying more PAV cycles, it aged faster than the fresh binder. Indeed, the difference in LT PG decreased from 4.6°C to 1.2°C after three PAV cycles. This trend can be observed in the HT PG of these two binders. To evaluate the effect of oil content on aging susceptibility of recycled binders, an artificial RAP binder was recycled with 16% oil. The HT and LT PG of the binder decreased from 76.7°C to 54.2°C and -22.7°C to -34.6°C, respectively. After applying four PAV cycles to the recycled binder, its LT increased from -34.6°C to -26.2°C. Fig 3 shows that the interpolated (FBP2+6%oil) which is the average oil content based on LT and HT PG showed acceptable performance in the long-term.

The aging rates for different types of binders are shown in Fig 4. Of the three oil contents tested (2.9%, 9.2% and 16%), recycled binder with 16% oil showed the least aging susceptibility and recycled binder with 2.9% oil showed the highest aging susceptibility to long-term aging. The fresh binder showed the lowest aging rate based on HT and LT PG. The higher aging susceptibility of recycled binder has been also demonstrated by Grilli et al. (2017). The long-term aging rates of recycled binders with different oil contents can be very different. The aging rate actually reduced as the oil content increased. When the aging rates of FB and FBP2+2.9% are compared, the change in the aging rate at LT PG was higher than that at HT PG. Fig 4 shows that the aging rate at LT increased from 0.125 to 0.25, while that at HT increased from 0.335 to 0.449. Generally, as the binders age, the HT PG improves at a higher rate, while LT PG degrades at a lower rate (at least 50% lower) (Fig 4). This behavior increased the continuous UTI as the binders aged (Fig 5). The FBP2+2.9% and FB, which had the highest and the lowest aging susceptibility showed the highest and lowest continuous UTI during aging. Fig 4 and 5 indicated that aging susceptibility decreased as the oil content increased. It is suggested that the higher oil content can be used in the recycled RAP binder without degrading short-term performance of asphalt, including moisture and rutting resistance (Cucalon et al., 2017).

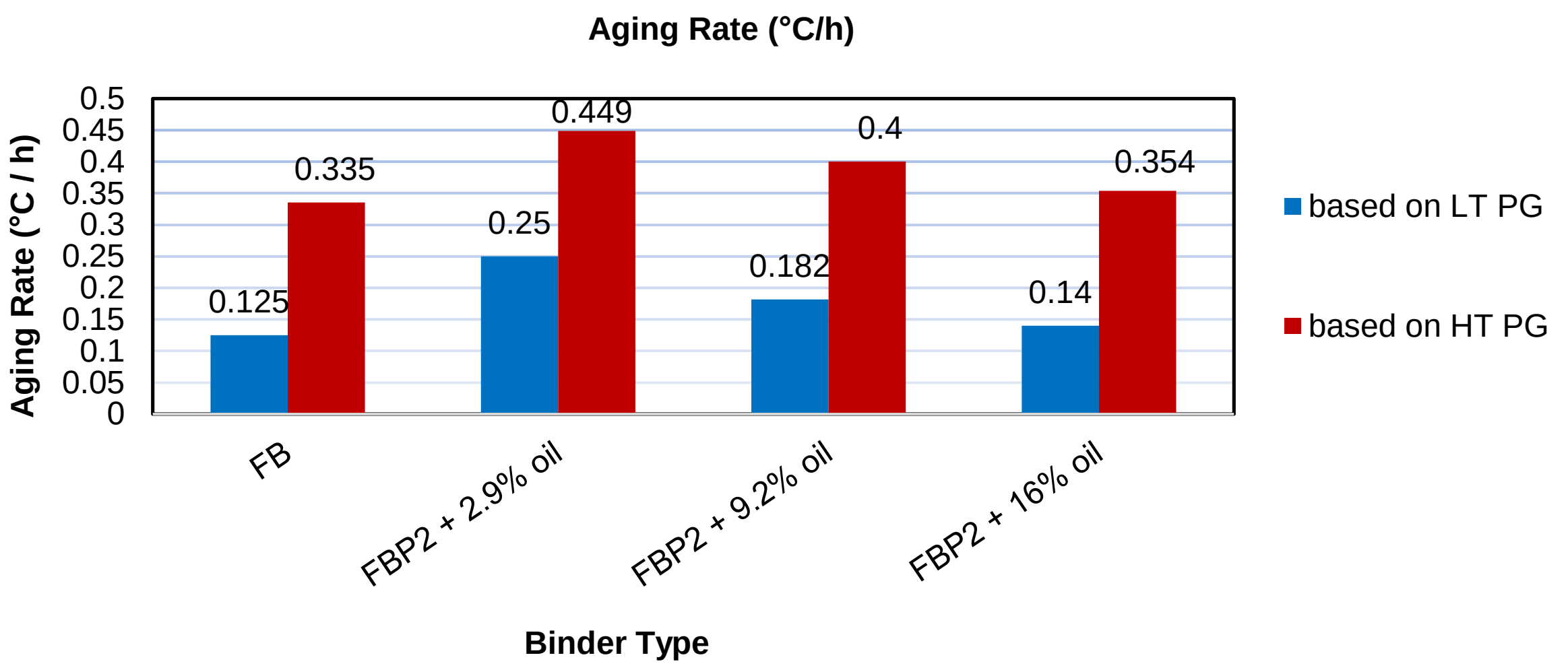

**Fig 4.** Aging rate of fresh and recycled binder based on LT and PG PG.

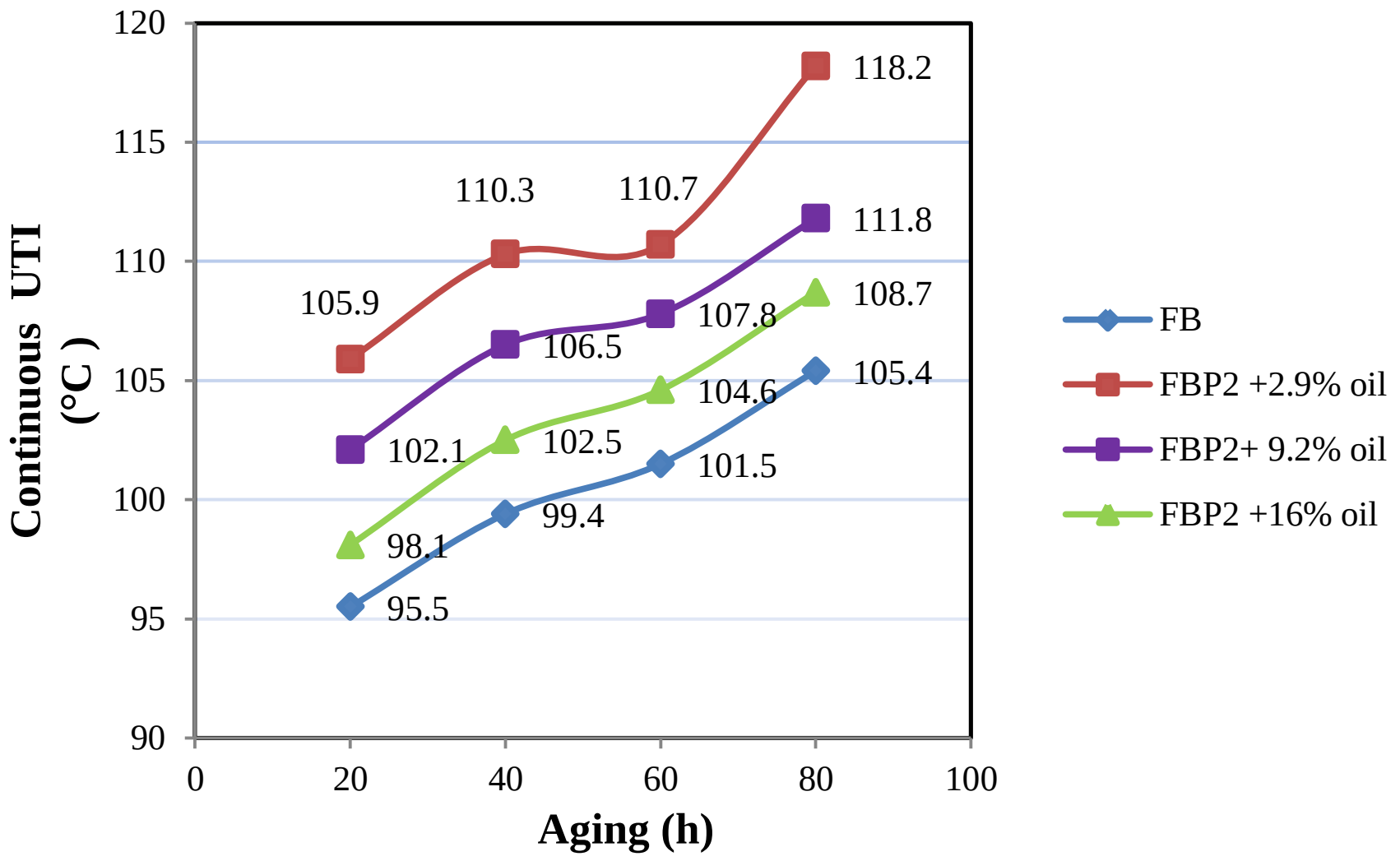

**Fig 5.** Increasing of continuous UTI during aging.

To examine the low-temperature cracking resistance of asphalt binders further, the continuous low-temperature grade based on stiffness and m-value at limiting values for 60 s were determined. Another factor used as a LT performance factor is $\Delta T_{cr}$ (Equation (2)). $\Delta T_{cr}$ represents the difference between the LT PG based on creep stiffness criteria (S = 300) and m-value (m = 0.300) (Anderson et al., 2011).

$$\Delta T_{cr} = Tcr\ (stiffness) - Tcr\ (m\text{-}slope) \quad (2)$$

Tcr (stiffness) is the critical low temperature where S(60) = 300 MPa,

Tcr (m-slope) is the critical low temperature where m(60) = 0.300.

Fig 6 shows that, for the two binders, the ΔTcr became more negative with increased aging. This was caused by the fact that aging decreased the m-value parameter faster than the change in stiffness of the binders. Furthermore, by comparing the FB and FBP2+2.9% oil curves (compare -6 and -9.8), the greater effect of oil on reducing the stiffness of the RAP binder than on increasing its m-value is demonstrated.

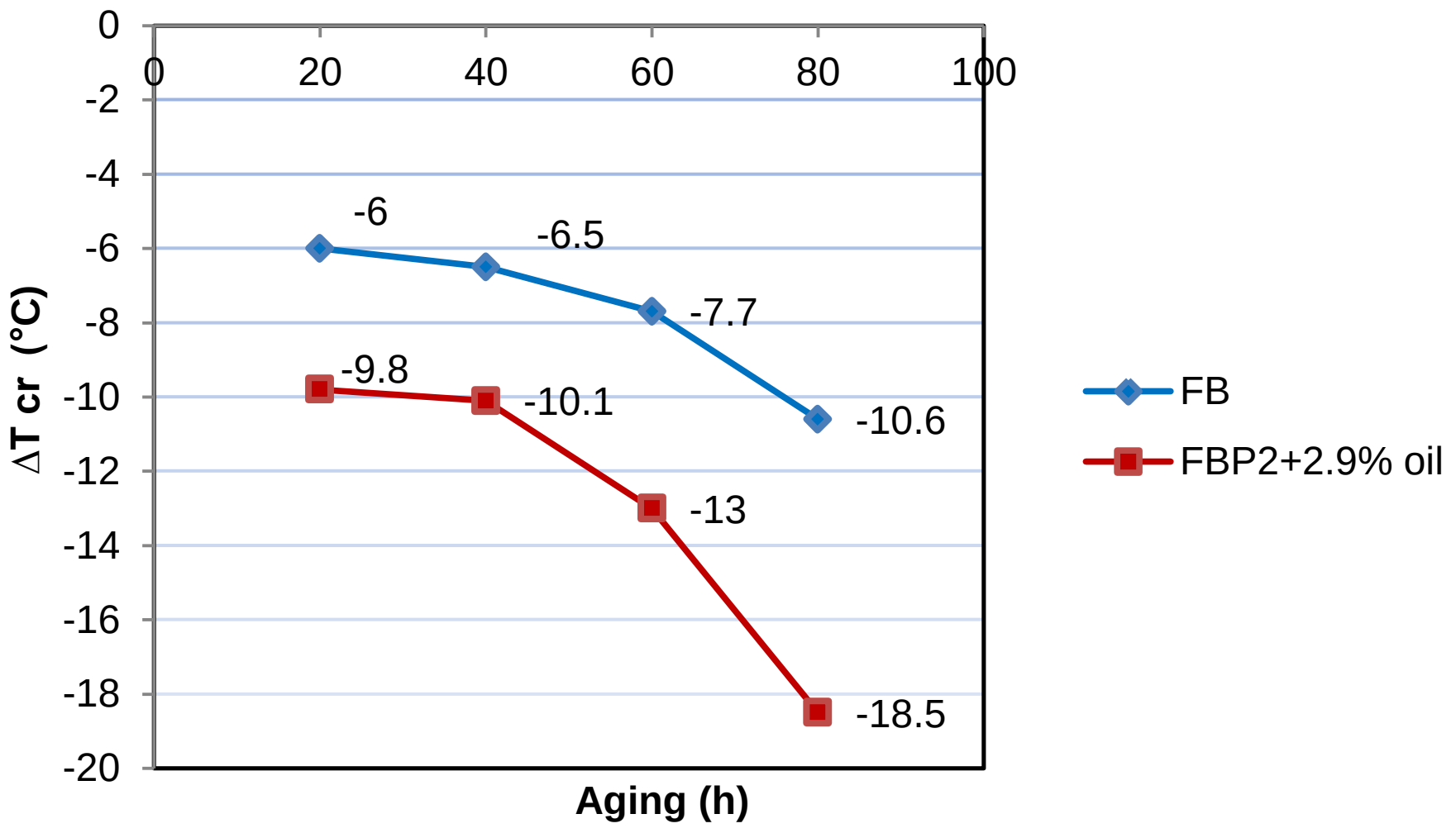

**Fig 6.** Reducing of $\Delta T_{cr}$ (°C) versus aging.

*4.2. Moisture susceptibility*

In this study, waste cooking oil was used to improve the RAP binder characteristics which consequently increased moisture susceptibility. Fig 7 shows that the four mixtures possessed the minimum TSR (80%) suggested by AASHTO T283.

The virgin mixture had the highest moisture susceptibility because of the high moisture susceptibility of the virgin, silica aggregates. The TSR parameter is the ratio of wet-to-dry indirect tensile strength (ITS). It is evident that additives affect both the dry and wet strengths. The addition of oil decreased the dry and wet strengths; thus, the TSR parameter cannot fully explain the effect of the oil. To compare the results, it is suggested that the ITS of wet specimens should be considered as well as their TSR ratio. Fig 7 indicates that the 60%RAP+CRM mixture had the largest wet ITS and the 60%RAP+FB+16%O had the smallest. The 100%RAP+FB+10%O mixture showed the best performance for moisture susceptibility among the mixtures which meet the volumetric specifications.

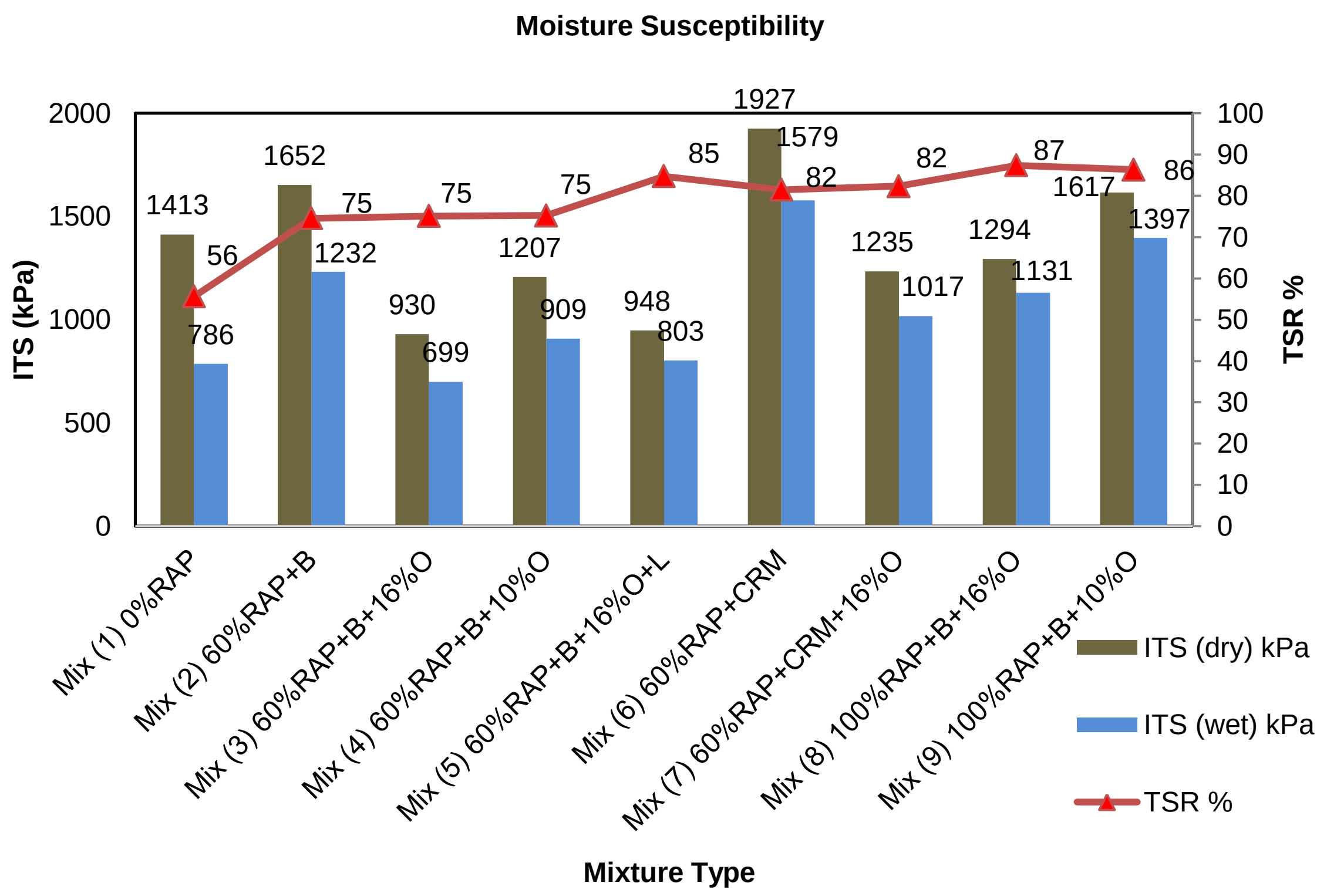

**Fig 7.** Moisture susceptibility of mixtures.

Fig 7 revealed the following results:

- RAP effect: The ITS wet and TSR increased 56.8% and 33.9%, respectively, and consequently moisture resistance increased (difference between mix (1) and mix (2)).
- Oil effect: The ITS wet decreased 43.3%, which is the lowest ITS wet from among the mixtures (difference between mix (2) and mix (3)). Recall that lower mixture stiffness is usually associated with lower ITS. The 16% oil was not suitable as indicated by the noticeable reduction in moisture resistance. However, If TSR is solely used to compare the moisture resistance of mixtures (equal value), a passing result will be obtained.
- A decrease in oil content increased the ITS wet (moisture resistance; difference between mix (3) and mix (4)).
- Lime effect: The ITS wet and TSR increased 14.9% and 12.8%, respectively, which increased the moisture resistance (difference between mix (3) and mix (5)).
- Crumb rubber effect: The ITS wet and TSR increased 28.1% and 9.2%, respectively, which increased the moisture resistance (difference between mix (2) and mix (6)).

- Effect of oil on crumb rubber mixture: ITS wet decreased 35.5%. This result demonstrated that oil reduced the moisture resistance (difference between mix (6) and mix (7)). Again, if the TSR factor is solely used to evaluate the moisture resistance of mixtures (equal value), a passing will be obtained and the greatly reduced tensile strength effect will be masked.

As the RAP aggregates are already coated with asphalt binder, high-RAP mixtures are generally less susceptible to stripping than virgin asphalt mixtures (Karlsson and Isacsson, 2006; Tran et al., 2012). The higher stiffness in RAP mixtures is also usually correlated with higher indirect tensile strength. Generally, moisture susceptibility is dependent on the adhesive resistance between the aggregate and binder, and the cohesive resistance in the binder mastic. The level of adhesive resistance in the RAP mixtures was greater than in the virgin mixes. The ITS wet increased in the 60%RAP+FB mixture compared to virgin one. The addition of oil to the RAP mixture decreased the wet ITS significantly. This mixture had the highest adhesive resistance, but an expected decrease in cohesive resistance was caused by the addition of oil oil (Al-Qadi et al., 2014), which led to the lowest ITS wet of all mixtures. The higher moisture susceptibility of the oil modified mixtures can be explained by the hydrophilic molecular structure of the waste vegetable oil, which increased the solubility of the oil-modified binder (Gong et al., 2015).

Interpolation of the ITS content for 13% oil in high-RAP mixtures show that an average content of the 10% and 16% oil (13% oil) will provide appropriate moisture resistance.

## 5 Conclusions

The use of waste cooking oil as a recycling agent opens the possibility for the routine design of high-RAP mixture. Although the waste oil was found to improve workability and low temperature performance, it reduced moisture and rutting resistance. It was found that the oil content should be carefully selected to create a balanced tradeoff between these characteristics. The blended binder PG with a high oil content that satisfied the HT PG criterion led to superior LT PG and acceptable HT PG in high-RAP mixtures (61.1-27 for 60%RAP). The results of long-term performance testing on recycled binders indicates that selecting the optimum oil content based on HT PG criteria can assure adequate performance of the recycled binder in the long run (better LT PG than the fresh binder after aging).

The results of moisture tests and HT PG indicate that, to the contrary, determining the oil content based on HT PG did not assure the performance of high-RAP mixtures in the short-term. It is important that the recycled asphalt assures performance in the short-term, then considering the performance of asphalt in long-term can be logical. Selecting the optimum oil content as equal to the average oil content based on satisfying the LT and HT PG can assure short-term performance of pavement and desirable long-term performance can be achieved. The results of moisture testing on mixtures, and PG testing on the blended binder showed that 13% oil (average of oil content for actual RAP binder) provided acceptable performance. The results of susceptibility to aging on recycled binder showed that adding 6% oil (average of oil content for artificial RAP binder) led to appropriate aging resistance. The major findings from this study are as follows:

1. The BBR and DSR results indicate that oil reduced the LT and HT PG of blended binders. Low temperature cracking resistance increased while the rutting resistance decreased. Also, oil has a greater on reducing the stiffness of the RAP binder than on increasing its m-value.
2. Recycled binders with waste cooking oil aged faster than fresh binder in the long-term, although their LT PG after the first PAV cycle was the same as fresh binder. This indicates that selecting optimum oil content based on LT PG criteria (2.9%) cannot assure adequate recycled binder performance in the long-term. Aging rates during the PAV cycles decreased as the oil content increased. The aging results indicate that selecting the optimum oil content based on HT PG (9.2%) produced appropriate performance in the long run. This confirms the importance of investigating aging susceptibility of recycled binders in the long-term.
3. Generally, as the binders aged, the HT PG improved at a higher rate while the LT PG degraded at lower rate (at least 50% lower). This led an increase in continuous UTI as the binders aged.
4. $\Delta T_{cr}$ decreased during aging and indicates that aging decreased the m-value parameter more than the stiffness of the binders. Comparison of the FB and FBP2+2.9% oil curves demonstrates the greater effect of oil on reducing the stiffness of RAP binder as compared to increasing the m-value.
5. In high-RAP mixtures, IDT strength (AASHTO T283) increased in comparison with the virgin mixture. This parameter decreased dramatically after adding oil to the RAP mixtures and is related to the decrease in cohesive resistance in the mastic.

6. Crumb rubber improved the moisture resistance of RAP mixtures. The use of a recycling agent along with crumb rubber in high-RAP mixtures increased the workability and compactability of the mixtures investigated. Using crumb rubber and recycling agents in high-RAP mixtures produced a high useful temperature interval (UTI) and appropriate moisture resistance. This result confirmed the use of crumb rubber with a recycling agent in high-RAP mixtures.
7. The results suggest the possibility of using high RAP contents of 60% and 100% in mixtures considering the suitable content of recycling agents and additives.

It is recommended to consider the road climate and condition to select the optimum recycling agent content. For hot climates, the optimum recycling agent content should be selected to satisfy the LT PG. In cold climates, a higher content that satisfies the HT PG should be selected. It is also suggested to determine and use the minimum appropriate level of recycling agent for mixture production to assure the short-term performance of the pavement. With time, a recycling agent can be sprayed onto the pavement during its service life to maintain the long-term performance characteristics of the pavement surface. More research is needed to investigate the optimum oil content in high-RAP mixtures and their performance in the long run.

**Conflict of interest**

The authors do not have any conflict of interest with other entities or researchers.

**Acknowledgments**

The authors would like to thank the Asphalt Pavement Lab at the Sharif University of Technology for providing the experimental equipment for testing.